\documentclass[11pt]{article}

\usepackage{jheppub} 
                     
\usepackage{braket,slashed,bm}
\usepackage{array,multirow}
\usepackage[normalem]{ulem}
\usepackage[T1]{fontenc} 

\usepackage{mathrsfs}

\usepackage{tikz}
\usetikzlibrary{arrows,decorations.pathmorphing,backgrounds,positioning,fit,petri,automata,shadows,calendar,mindmap,
decorations.markings,calc}

\def\rd{{\rm d}}

\def\lsix{ \mathcal{L}^{(6)}}

\def\hyp{\mathsf{y}}

\def\lsix{ \mathcal{L}^{(6)}}
\def\gcb{{\overline g_{1}}}
\def\gcw{{\overline g_{2}}}

\def\tc{{\overline \theta}}

\def\sc{{\overline s}}

\newcommand{\be}{\begin{equation}}
\newcommand{\ee}{\end{equation}}
\newcommand{\bea}{\begin{eqnarray}}
\newcommand{\eea}{\end{eqnarray}}
\newcommand{\ba}{\begin{array}}
\newcommand{\ea}{\end{array}}

\newcommand{\nn}{\nonumber}

\preprint{CERN-PH-TH/2014-187}

\title{On the consistent use of Constructed Observables.}

\author{
Michael Trott$^{1,2}$\\
$1$, Theory Division, Physics Department, CERN, CH-1211 Geneva 23, Switzerland,\\
$2$, Niels Bohr International Academy,
University of Copenhagen, Blegdamsvej 17, DK-2100 Copenhagen, Denmark}

\abstract{
We define "constructed observables" as relating experimental measurements to terms in a Lagrangian
while simultaneously making assumptions about possible deviations from the Standard Model (SM), in other Lagrangian terms. 
Ensuring that the SM effective field theory (EFT) is constrained correctly when using 
constructed observables requires that their defining conditions are imposed on the EFT in a manner that is consistent with the equations of motion.
Failing to do so can result in a "functionally redundant" operator
basis\footnote{We define the concept of functional redundancy, which is distinct from the usual concept of an operator basis redundancy, in the introduction.} and the wrong expectation as to how experimental quantities are related in the EFT. We illustrate the issues 
involved considering the $\rm S$ parameter and the off shell triple gauge coupling (TGC)
verticies. We show that the relationships between
$h \rightarrow V \bar{f} \, f$ decay and the off shell TGC verticies are subject to these subtleties, 
and how the connections between these observables vanish in the 
limit of strong bounds due to LEP. 
The challenge of using constructed observables to consistently constrain the Standard Model EFT is only expected to
grow with future LHC data, as more complex processes are studied.} 

\begin{document}
\maketitle

\section{Introduction}

Run one at LHC discovered a Higgs-like boson, and 
beyond the Standard Model (BSM) particles where not discovered, for masses $\lesssim 1 \, { \rm TeV}$. 
This has led to interest in effective field theory (EFT) 
approaches to Standard Model (SM) processes.
In this paper we discuss a subtlety that is present when constraining higher dimensional operators in an EFT, using an operator basis reduced by the Equations of Motion (EoM).

We will illustrate this point with the Standard Model effective field theory (SMEFT), which assumes that $\rm SU(2) \times U(1)_Y$ is linearly realized in the scalar sector, and that this symmetry is spontaneously broken by the SM Higgs. The dimension six operators are suppressed by $1/\Lambda^2$. LHC results indicate 
$\Lambda \gg v= 246\,\text{GeV}$, which provides a straightforward EFT expansion. The minimal classification of higher dimensional operators for this theory was given in
Ref.~\cite{Grzadkowski:2010es}, which further reduced the operator basis of a previous classification \cite{Buchmuller:1985jz}, by the classical EoM for the SM fields.  Although the reduction of the basis is a useful step, when considering experimental constraints on the reduced basis, subtleties can appear. 

Here we discuss one such subtlety. $S$ matrix elements correspond to physical quantities, but Wilson coefficients in a Lagrangian can be unphysical. 
The EoM relate different operators, with completely different field content 
in some cases, and yet $S$ matrix elements are unchanged by the EOM. One can remove an
operator entirely from a basis, and yet the physical effects present in the theory are not changed,
as $S$ matrix elements are not changed by the EOM. In this manner, the invariance of field theories under field redefinitions \cite{Weinberg:1968de,Callan:1969sn,Politzer:1980me}
shows that an operator basis is unphysical.  

At the same time, when constraining the SMEFT at the Lagrangian level, there is a conservation of
constraints in changing basis. 
The subtlety discussed in this paper corresponds to the case when observables are constructed from the data to determine such constraints, with a series of assumptions imposed about the nature of possible deviations in the SMEFT, i.e under the assumptions that certain parts of Feynman diagrams are as in the SM. These defining conditions can introduce subtle constraints onto
the field theory.\footnote{We avoid the use of the phrase pseudo-observable in this paper, due to its various historical definitions in the literature, see Ref \cite{Gonzalez-Alonso:2014eva} for a recent comprehensive discussion on pseudo-observables in the SMEFT.}The theory can be properly constrained
if the defining conditions of the observables are incorporated in a basis independent manner, in conjunction
with the constructed experimental bound. Failing to do so
can lead to a {\it functionally redundant} operator basis, in that the number of parameters present in the Lagrangian is inconsistent with the assumptions required
to incorporate the bound from a constructed observable.\footnote{It is important to distinguish the standard definition of operator redundancy, where a basis is being used that is not reduced fully with the EoM,
from a functional redundancy. The latter can still occur in a fully reduced basis if constructed observables are used in an inconsistent manner. The confusion 
that results from either redundancy is the same.}  A concrete example of a functional redundancy is given in Fig 1.

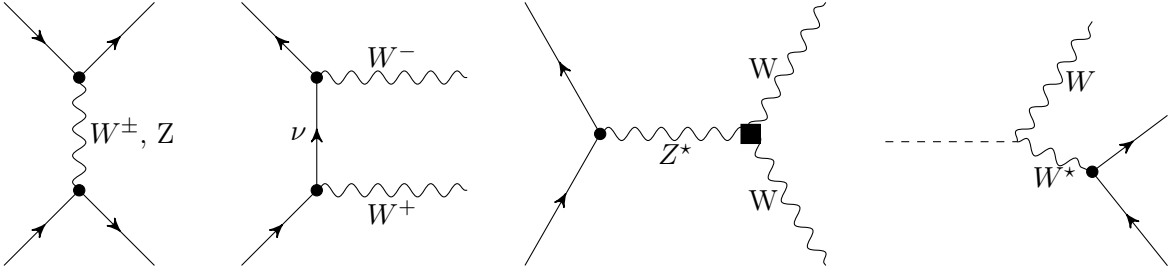
\begin{figure}
\begin{tikzpicture}[
decoration={
	markings,
	mark=at position 0.55 with {\arrow[scale=1.5]{stealth'}};
}]
\draw[postaction=decorate]  (0,0) --(1,1) ;
\draw[postaction=decorate]  (1,1) --(2,0) ;
\filldraw (1,1) circle (0.075);
\draw[decorate,decoration={snake}] (1,1) --   node [right] {$W^{\pm}$, Z} (1,2.5);
\filldraw (1,2.5) circle (0.075);
\draw[postaction=decorate]  (1,2.5) -- (2,3.5) ;
\draw[postaction=decorate]   (0,3.5) -- (1,2.5) ;
\end{tikzpicture}
\hspace{0.5cm}
\begin{tikzpicture}[
decoration={
	markings,
	mark=at position 0.55 with {\arrow[scale=1.5]{stealth'}};
}]
\draw[postaction=decorate]  (0,0) --(1,1) ;
\draw[postaction=decorate]  (1,1) -- node [left] {$\nu$}(1,2.5) ;
\draw[postaction=decorate]  (1,2.5) --(0,3.5) ;
\draw[decorate,decoration={snake}] (1,1) --   node [below] {$W^+$} (3,1);
\filldraw (1,1) circle (0.075);
\filldraw  (1,2.5) circle (0.075);
\draw[decorate,decoration={snake}] (1,2.5) --   node [above] {$W^-$} (3,2.5);
\end{tikzpicture}
\hspace{0.5cm}
\begin{tikzpicture}[
decoration={
	markings,
	mark=at position 0.55 with {\arrow[scale=1.5]{stealth'}};
}]
\draw[postaction=decorate]  (0,0) --(1,1.75) ;
\draw[postaction=decorate]  (1,1.75) --(0,3.5) ;
\filldraw (1,1.75)  circle (0.075);
\draw[decorate,decoration={snake}] (1,1.75) --   node [below] {$Z^\star$} (3,1.75)
node [rectangle,draw=black,fill=black] {};
\draw[decorate,decoration={snake}] (3,1.75) --   node [left] {W} (4,3.5);
\draw[decorate,decoration={snake}] (3,1.75) --   node [left] {W} (4,0);
\end{tikzpicture}
\hspace{0.5cm}
\begin{tikzpicture}[
decoration={
	markings,
	mark=at position 0.55 with {\arrow[scale=1.5]{stealth'}};
}]
\draw [dashed] (0,2.4) --(1.75, 2.4) ;
\draw[decorate,decoration={snake}] (1.75, 2.4) --  node [below] {$W^\star$} (2.75, 2);
\draw[decorate,decoration={snake}] (1.75, 2.4) --  node [right] {$W$}  (2.75,4);
\draw[postaction=decorate]  (2.75, 2) --(3.75,2.75) ;
\draw[postaction=decorate]  (3.75,0.75) --(2.75, 2) ;
\filldraw (2.75, 2)  circle (0.075);
\end{tikzpicture}
\caption{\label{fig:1} 
An example of a functional redundancy. An operator basis can be chosen that maps parameters characterizing
differences in the coupling of the $W$ and $Z$ to leptons (compared to the SM) into another sector of the field theory, where these parameters contribute to an anomalous TGC vertex.
(Parameters can be mapped from the dot in the diagrams above to the box with the EoM.)
Subsequently, using a TGC vertex bound, naively constrains these parameters in the SMEFT. Some of the parameters apparently constrained in this manner
are functionally redundant,  as in the middle two diagrams the production and decay of the $W,Z$ is simultaneously assumed to be SM like.
(When experimental bounds are constructed on the parameters in the box, the dot is assumed to be SM like.) 
This procedure is inconsistent and does not constrain a flat direction due to LEP $Z$ pole data that can modify $h \rightarrow V \mathcal{F}$ decay,
when $V = W$. Unphysical field redefinitions, or an operator basis choice, do not make this procedure consistent.}
\end{figure}

We illustrate the basic issues involved in Section \ref{defn}.
Constraints due to LEP data on the SMEFT are discussed in Section \ref{LEPdata}.
The impact of the defining conditions for the oblique electroweak precision data (EWPD) $\rm S$ parameter, and the off shell TGC verticies are discussed in Section \ref{pseudoobservablesection}.
We then show that reporting the relationship between the differential spectra in $h \rightarrow V \bar{f} \, f$ decay\footnote{In this paper, we will  consistently use the notation $V$ for $W,Z$, a general massive gauge boson.}
and off shell TGC verticies has a potential basis dependence due to this issue, and how to resolve this problem 
by taking into account constraints of this form in Section \ref{hvfspec}. We find that in the limit of strong constraints from LEP data (we define this limit precisely below), off shell TGC verticies are not related to  $h \rightarrow V \bar{f} \, f$ decay spectra. 

Our results make clear that data analyses can benefit from using (at least) two bases, with careful attention paid to the EoM mapping between them.\footnote{One might consider 
it even more ideal to have no basis at all. However, combining constraints from multiple scales, and correlating such information with future higher energy measurements
requires the machinery of perturbative corrections.}
The subtlety discussed here is relevant to future efforts to obtain more precise constraints, from more complex final state studies at LHC.
In analyzing such processes, constructed observables will be extracted if simplifying assumptions that do not generate Ward identities are made about the nature of possible deviations from the SM.
\section{Operator relations due to the EoM}\label{defn}

We adopt notation for the linear SMEFT consistent with Ref \cite{Grojean:2013kd,Jenkins:2013zja,Jenkins:2013wua,Alonso:2013hga,Alonso:2014rga,Alonso:2014zka}.(With the shorthand $\bar{s}_\theta = \sin \bar{\theta},\bar{c}_\theta = \cos \bar{\theta}$. The notation is also summarized in the Appendix.) 
The Lagrangian  $ \lsix = \sum_i C_i \, Q_i$
consists of all dimension six operators that can be constructed preserving $\rm SU(3)_C \times SU(2)_L \times U(1)_Y$ (linearly), and assuming the conservation of baryon and lepton number. 
Taking into account flavour indicies, there are $2499$ parameters to constrain in $\lsix$, as shown in  Ref.\cite{Alonso:2013hga}. Despite this large number, the EoM have been used extensively to reduce the number of parameters to a minimal set. The SM EoM are summarized in the Appendix. 

It is well known that a choice of operator basis is arbitrary and cannot effect a physical conclusion, such as how strongly constrained an EFT is by an experimental measurement.
Considering the EoM makes clear the requirement of thinking of a Wilson coefficient as an ensemble parameter that can obtain experimental constraints from all possible measurements that can constrain the parameter in any basis.(So long as measurements are not reused.) The EoM can also make clear the consequences of defining conditions for constructed observables.
Careful use of the EoM is the easiest way to avoid a functional redundancy. 
A simple example of the ensemble nature of the Wilson coefficient, and how the EoM can be useful, is afforded with the dimension six operator
\begin{align}
E_{H \Box} &= [H^\dagger H] [H^\dagger (D^2 H) + (D^2 H^\dagger) H],
\label{ebox}
\end{align}
this operator can be converted via Eqn \ref{eomHiggs} to 
\begin{align}
\widetilde E_{H \Box} & = 2\lambda v^2 (H^\dagger H)^2 -4 \lambda Q_H -\left( [Y_u^\dagger]_{rs}\, Q_{\substack{u H \\ rs }} + [Y_d^\dagger]_{rs} \, Q_{\substack{d H \\ rs }}+ [Y_e^\dagger]_{rs}\, Q_{\substack{eH \\ rs }}+ \hbox{h.c.} \right).
\label{ebox2}
\end{align}
Note here the introduction of the operators $Q_{\substack{u H \\ rs }}$ etc, which are matricies in flavour space in general, with flavour indicies $r,s$ contracted with the SM
Yukawa matricies. The SM Yukawa matricies are defined in the Appendix. 
Define $\mathcal{R} =  \lsix  + C_{E \Box} \, E_{H \Box}$ and
applying Eqn \ref{ebox2} to reduce $\mathcal{R}$ to $ \lsix$ gives the following parameter redefinitions at a chosen scale
\begin{align}
C'_H & = C_H - 4 \, \lambda C_{E\Box}, &
C'_{\substack{u H \\ rs }} & =  C_{\substack{u H \\ rs }}  - C_{E\Box} \, [Y_u^\dagger]_{rs}, \nn \\
C'_{\substack{d H \\ rs }} & =  C_{\substack{d H \\ rs }}  - C_{E\Box} \, [Y_d^\dagger]_{rs}, &
C'_{\substack{e H \\ rs }} & =  C_{\substack{e H \\ rs }}  - C_{E\Box} \, [Y_e^\dagger]_{rs}, \nn \\
\lambda' & =  \lambda + 2 \lambda v^2\, C_{E\Box}.
\label{parametershift}
\end{align}
The hermetian conjugate Wilson coefficients of $C_{\substack{u H}},C_{\substack{d H}},C_{\substack{e H}}$ are similarly 
shifted. Now consider two bases. In the first, one choses to remove $Q_H$ in favour of $E_{H \Box}$, in the second one choses to remove $E_{H \Box}$ in favour of $Q_H$. 
The Wilson coefficients are identified when changing basis in this case:
$C_{E\Box} \equiv - 4 \, \lambda C_H$. 
The same parameter in the field theory can obtain direct constraints from
measurements that constrain $C_{H \Box}$ in basis one, and $C_H$ in basis two, even though the field content present in the operators differ. 
The constraints obtained in the two bases are related by the EoM, and the strongest constraint is relevant for the optimal basis independent bound on the EFT.
A functional redundancy would be present if the parameter $C_{E\Box}$ is retained, while {\it simultaneously} a constructed observable
was used to constrain the field theory that assumed $C_H = 0$.

This point holds for more complicated basis changes. Two bases of operators are of interest in the following sections, the basis of Ref.~\cite{Grzadkowski:2010es}, and the basis used in Ref.~\cite{Pomarol:2013zra}.
The former will also be referred to as the standard basis.\footnote{We emphasize that any basis is allowed, and no basis is superior to any other, if calculations are performed correctly
and functional redundancy is avoided. We refer to the standard basis here as
this basis was the first dimension six operator basis fully reduced by the SM EoM.} We will denote the operators in the later
basis with $\mathcal{O}$ labels to avoid confusion.  Define the Wilson coefficients to be 
\begin{align}\label{lagrangianequivalence}
\lsix &= \sum_i C_i \, Q_i = \sum_i \, \mathcal{P}_i \, \mathcal{O}_i.
\end{align}
$C_i$ and $\mathcal{P}_i$ have mass dimension $- 2$.
The operators that are present in the $Q_i$ but not the $\mathcal{O}_i$ are given
by
\begin{align}
Q_{HW} &= H^\dagger H \, W^{I}_{\mu\nu} \,  W_{I}^{\mu\nu}, &
Q_{\substack{H\ell \\pr}}^{(1)} &= (H^\dagger \, i\overleftrightarrow D_\mu H) \, \bar{\ell}_p \, \gamma^\mu \, \ell_r, &
Q_{HWB} &= H^\dagger \, \tau_I \, H \, W^{I}_{\mu\nu} \,  B^{\mu\nu}, \nn \\
Q_{\substack{H\ell \\pr}}^{(3)} &= (H^\dagger \, i\overleftrightarrow D^I_\mu H) \, \bar{\ell}_p \, \tau^I \, \gamma^\mu \, \ell_r, &
Q_{HD} &= (H^\dagger D^\mu H)^\star \, (H^\dagger D_\mu H).
\label{standardops}
\end{align}
The operators that are present in $\mathcal{O}_i$ and not in the $Q_i$ are given by
\begin{align}
\mathcal{O}_{HW} &=  -i \, g_2 \, (D^\mu H)^\dagger \, \tau^I \, (D^\nu H) \,  W^I_{\mu \, \nu}, & \hspace{1cm}
\mathcal{O}_{HB} &=  -i \, g_1 \, (D^\mu H)^\dagger \,  (D^\nu H) \,  B_{\mu \, \nu},\nn \\
\mathcal{O}_{W} &=  -\frac{i \, g_2}{2} \, (H^\dagger \,  \overleftrightarrow D^I_\mu H) \,  (D^\nu W^I_{\mu \, \nu}), & \hspace{1cm}
\mathcal{O}_{B} &=  -\frac{i \, g_1}{2} \, (H^\dagger \, \overleftrightarrow{D}^\mu H) \,  (D^\nu B_{\mu \, \nu}), \nn \\
\mathcal{O}_{T} &=  (H^\dagger \, \overleftrightarrow{D}^\mu H) \,  (H^\dagger \, \overleftrightarrow{D}^\mu H).
\label{SILHops}
\end{align}
The relevant relationships between the operators in these basis are completely given in Ref. \cite{Alonso:2013hga} (see Appendix B).\footnote{
Operator relations of this form were partially discussed in Refs \cite{SanchezColon:1998xg,Kilian:2003xt,
Grojean:2006nn} and many other works previously.} The  transformation from the standard basis
to the $\mathcal{O}_i$ basis is derived using the SM EoM\footnote{In these relations only the flavour singlet component of the operators appears. This is indicated with the notation $Q_{\substack{H d \\ rr}}$ for example,
which explicitly corresponds to an operator that is proportional to a unit matrix in flavour space.
}, and found to be
\begin{align}
g_1 \, g_2 \, Q_{HWB} &=  4 \, \mathcal{O}_{B} - 4 \, \mathcal{O}_{HB} - 2 \,  \hyp_H \, g_1^2 \, Q_{HB}, \nn \\
g_2^2 \, Q_{HW} &=  4 \, \mathcal{O}_{W} - 4 \, \mathcal{O}_{B} - 4 \, \mathcal{O}_{HW} + 4 \, \mathcal{O}_{HB} + 2 \,  \hyp_H \, g_1^2 \, Q_{HB}, \nn \\
g_1^2 \, \hyp_\ell \, Q_{\substack{H l \\ tt}}^{(1)} &= 2 \, \mathcal{O}_{B} +  \hyp_H \, g_1^2\, \mathcal{O}_T - 
g_1^2\left[\hyp_e Q_{\substack{H e \\ rr}} + \hyp_q Q_{\substack{H q \\ rr}}^{(1)}+\hyp_u Q_{\substack{H u \\ rr}}+ \hyp_d Q_{\substack{H d \\ rr}} \right], \nn \\
g_2^2  \, Q_{\substack{H l \\ tt}}^{(3)} &=  4 \, \mathcal{O}_{W} - 3 \, g_2^2 \, Q_{H \Box} + 2 \, g_2^2 m_h^2 \, (H^\dagger \, H)^2 - 8 \, g_2^2 \, \lambda \, Q_H - g_2^2 \, Q_{H q}^{(3)}, \nn \\ 
&- 2 \, g_2^2\left( [Y_u^\dagger]_{rr} Q_{\substack{ uH \\ rr}} + [Y_d^\dagger]_{rr} Q_{\substack{dH \\ rr}}+ [Y_e^\dagger]_{rr} Q_{\substack{eH \\ rr}}+h.c. \right).
\label{inversetransform1}
\end{align}
Some parameters are only redefined changing basis, 
and a constraint is lost in the arbitrariness of redefining parameters.
This is not always the case. Considering the case of interest, we find the mapping
\begin{align}\label{mappingparameters}
\mathcal{P}_B &\rightarrow \frac{4}{g_1 \, g_2} C_{HWB} - \frac{4}{g_2^2} \, C_{HW} + \frac{2}{g_1^2 \, \hyp_\ell} \, C_{\substack{H \ell \\tt}}^{(1)},  \quad \quad \quad  \quad \mathcal{P}_W \rightarrow \frac{4}{g_2^2} \, C_{HW} + \frac{4}{g_2^2} \, \, C_{\substack{H \ell \\ tt}}^{(3)}, \nn \\
\mathcal{P}_{HB} &\rightarrow -\frac{4}{g_1 \ g_2} \, C_{HWB} + \frac{4}{g_2^2} \, \, C_{HW}, \hspace{2.85cm}
\mathcal{P}_{HW} \rightarrow -\frac{4}{g_2^2} \, C_{HW}.
\end{align}
This mapping is obtained by using Eqn \ref{inversetransform1} in Eqn \ref{lagrangianequivalence}.
These parameters in the $\mathcal{O}$ basis are identified with alternate parameters in the standard basis.\footnote{This is true at a fixed scale, when the RGE evolution of the theory is taken
into account, this relationship will be weakened by loop corrections.} The choice that has been made in constructing this basis is to remove operators directly related to $V$ decay and phenomenology, and to map possible differences in $Z$ and $W$ couplings to leptons in the SMEFT to a different sector of the field theory. When strong constraints on the parameters $C_{\substack{H \ell \\tt}}^{(1)},C_{\substack{H \ell \\tt}}^{(3)},C_{HWB}$ are present, this results in a large degree of non intuitive hidden correlations in the $\mathcal{P}_i$ Wilson coefficients. Of course the converse is also true, constraints on the $\mathcal{P}_i$ lead to 
non intuitive hidden correlations on the $C_i$ Wilson coefficients. There is no intrinsically intuitive basis, as a basis choice is unphysical.

It is well known that setting an operator to zero for a measurement, and removing the same operator
with the EoM are not equivalent procedures. A consequence of this fact is that using field redefinitions to attempt to satisfy the defining condition of a constructed observable
corresponds to a poor choice of basis.
A defining condition is still present for the constructed observable in this case, consistency requires this {\it always} leads to a constraint on the field theory. 
The constraint will simply be non intuitive and the resulting basis can be functionally redundant.
Another important consequence of this fact is that removing parameters by field redefinitions,
as they are considered to be strongly experimentally bounded and irrelevant for future experimental studies, can also lead to a functionally redundant basis. 
Using field redefinitions in this manner is in general a mistake.

\section{LEP data}\label{LEPdata}
The discussion of the previous section is relevant to efforts to constrain the SMEFT with LHC and pre-LHC (LEP, Tevatron and other EW) data.
Considering pre-LHC data, we will take as input parameters the measured values
of the fine structure constant $\hat{\alpha}_{ew}$ (from the low energy limit of electron Compton scattering), the fermi decay constant in muon decays $\hat{G}_F$ and the measured $Z$ mass ($\hat{m}_Z$). It is convenient to relate observables in terms of the parameters $g_2, \sin^2 \theta = g_1^2/(g_1^2 + g_2^2)$ and the electroweak vev {\it{v}}.
Defining at tree level the effective {\it measured} mixing angle
\bea\label{sinequation}
\sin^2 \hat{\theta} = \frac{1}{2} - \frac{1}{2}\sqrt{1 - \frac{4 \, \pi \hat{\alpha}_{ew}}{\sqrt{2} \, \hat{G}_F \, \hat{m}_Z^2}},
\eea
then the measured value of a gauge coupling can be inferred as
\bea
 \hat{g}_2 \, \sin \hat{\theta} = 2 \, \sqrt{\pi} \, \hat{\alpha}_{ew}^{1/2}.
\eea
The measured vev can be defined as $\hat{v}^2 = 1/\sqrt{2} \, \hat{G}_F$.

\subsection{Parameter counting and LEP data}
The number of parameters present to constrain in the lepton sector are two parameters corresponding to $C_{HWB},C_{HD}$, $(n_g^2 + n_g^4)/2$ parameters for the coefficient 
 $C_{\substack{ll \\ prst}}$ with $n_g=3$ generations of leptons, and $n_g^2$ parameters for each of $C^{(3)}_{\substack{Hl \\ pr}},C^{(1)}_{\substack{Hl \\ pr}},C_{He}$. Finally, the Wilson coefficient of the operator $(\bar{e}_p \gamma_{\mu} e_r)(\bar{e}_s \gamma^{\mu} e_t)$ corresponds to $n_g^2 (1+ n_g)^2/4$ parameters.  The total number of parameters sums to 110 in the lepton sector in the standard basis.

In the $\mathcal{O}$ basis three of these parameters:  $C_{HWB}, C^{(3)}_{\substack{Hl \\ tt}},C^{(1)}_{\substack{Hl \\ tt}}$, are chosen to be mapped to alternate parameters
using the EoM operator relations. $C_{HD}$ is exchanged for $\mathcal{P}_T$, and  the operator $Q_{HW}$ is exchanged for 
$O_{HW}$. This leads to a net reduction of two parameters $C^{(3)}_{\substack{Hl \\ tt}},C^{(1)}_{\substack{Hl \\ tt}}$ in some of the well
measured EWPD observables. 

To constrain $\lsix$, there are the lepton flavour specific
LEP observables $\mathcal{A}_\ell, R_\ell$, $\sigma_{had}^0, \Gamma_Z$, reported results on the $\rho$ parameter,
inferred constraints on the EWPD parameters from global fits, and TGC verticies.
EWPD and TGC verticies are not directly observable and are discussed in the following sections.
In both bases, there are not enough reported measurements to constrain all the parameters model independently.

As a result, simplifying assumptions are made. One can neglect the effects of some four fermion operators, assuming that there are no significant hierarchies
in the Wilson coefficients to counteract their relative $\Gamma_Z^2/M_Z^2 \sim 10^{-3}$ suppression, in this case 22 parameters are relevant.
Further neglecting parameters related to flavour violation reduces the number of parameters down to ten. 

A simplified scenario where all flavour structure in BSM physics is assumed to be vanishingly small is sometimes also considered.
This corresponds to adopting a strict  $\rm U(3)^5$ flavour symmetry assumption consistent with MFV \cite{DAmbrosio:2002ex} in the SMEFT. In this case, $n_g = 1$, and the number of free parameters is trivialized down to seven
in the standard basis. Flavour universality in the leptonic decays of the $V$ is the difference between the ten and seven parameters quoted. Further neglecting the $(\bar{e}_t \gamma_{\mu} e_t)(\bar{e}_t \gamma^{\mu} e_t)$ operator leaves 6 parameters to constrain with LEP data.

\subsection{Constraints due to LEP data}\label{constraints}
Predicting observables in the SMEFT,
each of the measured input parameters has been shifted from its theoretical value in the SM. This shift has been
absorbed into the measured value. To aid in simplifying results,\footnote{In the following discussion we largely follow the
analysis of Ref \cite{Pomarol:2013zra}.The main difference is the use of the standard basis, and considering the EOM relations in Eqn \ref{mappingparameters} when comparing results between bases.}, we introduce the parameters
\bea
\mathcal{S} = \frac{v_T^2 \, C_{HBW}}{\bar{g}_1 \, \bar{g}_2}, \quad \quad \mathcal{T} = \frac{1}{2} v_T^2 \, C_{HD}. 
\eea

To leading order in the standard basis, the input parameters are modified (compared to the usual definition of these parameters in the SM Lagrangian) by a shift given by
\bea\label{parameterdefinition1}
\frac{\delta \alpha_{ew}}{(\alpha_{ew})_{SM}} &=& - 2 \, (s^{SM}_\theta)^2 \, \bar{g}_2^2  \,  \mathcal{S}, \nn \\
\frac{\delta G_F}{(G_F)_{SM}} &=&  - \frac{v_T^2}{2}  \left(C_{\substack{ll \\ \mu ee \mu}} +  C_{\substack{ll \\ e \mu\mu e}}\right) + v_T^2 \left(C^{(3)}_{\substack{Hl \\ ee }} +  C^{(3)}_{\substack{Hl \\ \mu\mu }} \right),  \nn \\
\frac{\delta m_Z^2}{(m_Z^2)_{SM}} &=& \mathcal{T}  + 2 \, (s^{SM}_\theta)^2 \, \bar{g}_2^2  \,  \mathcal{S}.
\eea

Parameterizing deviations in LEP data for $\Gamma_Z^L \equiv Z \rightarrow \bar{\ell}_L \, \ell_L, \Gamma_Z^R \equiv Z \rightarrow \bar{\ell}_R \, \ell_R, \Gamma_Z^\nu \equiv Z \rightarrow \bar{\nu} \, \nu$, and $m_W$, one finds 
\bea\label{LEPdata1}
\frac{\delta\Gamma_Z^{L(t)}}{\Gamma_Z^L} &=&  \frac{1}{\bar{c}^2_{2 \, \theta}} \left(\mathcal{T} + \frac{\delta G_F}{(G_F)_{SM}} + 4 \bar{s}^2_\theta \, \bar{g}_2^2 \, \mathcal{S} \right) + \frac{2 \, v_T^2}{2 \, \bar{s}^2_\theta - 1} \left(C_{\substack{H \ell \\tt}}^{(1)} + C_{\substack{H \ell \\ tt}}^{(3)} \right), \\ \label{LEPdata2}
\frac{\delta\Gamma_Z^R}{\Gamma_Z^R} &=& - \frac{1}{\bar{c}_{2 \, \theta}} \left(\mathcal{T} + \frac{\delta G_F}{(G_F)_{SM}} + 2  \, \bar{g}_2^2 \, \mathcal{S} \right) - \frac{v_T^2 \, C_{He}}{\bar{s}^2_\theta},  \\ \label{LEPdata3}
\frac{\delta\Gamma_Z^\nu}{\Gamma_Z^L} &=&  \mathcal{T} + \frac{\delta G_F}{(G_F)_{SM}} + 2 \, v_T^2 \left(C_{\substack{H \ell \\tt}}^{(1)} - C_{\substack{H \ell \\ tt}}^{(3)} \right),  \\
\label{LEPdata4}
\frac{\delta m_W}{m_W} &=& \frac{1}{2 \, \bar{c}_{2 \, \theta}} \left( \bar{c}^2_{\theta} \mathcal{T} +  \bar{s}^2_{\theta} \left(\frac{\delta G_F}{(G_F)_{SM}} + 2 \bar{g}_2^2 \, \mathcal{S} \right) \right).
\eea
The introduction of two extra parameters compared to the $\mathcal{O}$ basis leads to two purely unconstrained
parameters.\footnote{Note that other directions in the operator parameter space can be numerically less constrained due to accidental approximate cancelations in Eqn \ref{LEPdata1}-\ref{LEPdata4}. We refer to pure flat directions to make this distinction clear. The following discussion is consistent with a careful examination of the results of in Ref \cite{Han:2004az,Grojean:2006nn}. Note the distinction between pure flat directions and approximate flat directions due to numerical accidents
is relevant in this comparison. The t-channel $\nu$ exchange contribution to $\sigma(e^+ \, e^- \rightarrow W^+ \, W^-)$,  was included in the fit in Ref \cite{Han:2004az}. This  consistency does not extend to some subsequent literature.}
Despite $C_{\substack{H \ell \\tt}}^{(1)}, C_{\substack{H \ell \\ tt}}^{(3)}$ being present compared to the  $\mathcal{O}$ basis 
 the pure flat directions do not have to involve these parameters. There is no special basis.

Consider the case of the multiple Wilson coefficients present in $\delta G_F$, or the relation
 between $\mathcal{T}, \delta G_F$ and $C_{\substack{H \ell \\ tt}}^{(3)}$ in $\delta\Gamma_Z^{\nu}$ in the limit of the flavour trivialized SMEFT\footnote{It is interesting to note the nontrivial effects of the $\rm U(3)^5$ symmetry on this choice, and the difference in the 10 vs 7 parameters present.
In the case where flavour structure is not trivialized, each of the $\delta\Gamma_Z^{L(t)}$ for $t =e, \mu, \tau$ has an individual shift in Eqn \ref{LEPdata1}.
Conversely, in $\delta G_F$ the flavour specific sum $C^{(3)}_{\substack{Hl \\ ee }} +  C^{(3)}_{\substack{Hl \\ \mu\mu }}$
appears. Flat directions in LEP data are sensitive to lepton flavour symmetry assumptions in this manner.}. One can always choose the accidental relation
\begin{align}\label{unconstrained}
2 \, C^{(3)}_{\substack{Hl}} &= C_{\substack{ll}}, & v_T^2 \, C_{\substack{H \ell}}^{(3)} &=  \frac{\mathcal{T}}{2} +  \frac{\delta G_F}{2 \, (G_F)_{SM}}.
\end{align}
With this choice the dependence on $C^{(3)}_{\substack{Hl}}$ and $\delta G_F$ is removed in Eqn \ref{LEPdata1}-\ref{LEPdata4}. 
One can consider the remaining parameters constrained to have fixed relationships due to LEP measurements,
and then the above relations represent chosen pure flat directions (in this case $v_T^2 \, \mathcal{C}_{He} = \mathcal{T}$). This choice is arbitrary, as is any other in a system of unconstrained equations. 
This choice is interesting to consider, when examining off-shell TGC vertex bounds, as in this case the coupling of the $W$ and $Z$ to leptons are physically allowed to differ.

The Wilson coefficient $C_{HWB}$ exactly canceling against the parameters $C_{\substack{H \ell \\tt}}^{(1)},C_{\substack{H \ell \\ tt}}^{(3)}$
which has been argued to be relevant to the definition of the $S$ parameter (see Section \ref{Sparameter}), need not correspond to a pure unconstrained direction. If two more measurements are made, all of the parameters appearing in the lepton sector are then constrained.

\subsection{Lifting flat directions through scale dependence}
LEP data is not blind to the pure unconstrained directions resulting from Eqn \ref{LEPdata1}-\ref{LEPdata4}, before considering TGC verticies, as the operators
are scale dependent quantities. The full renormalization of the dimension six operators in the
SMEFT (with nontrivial flavour structure) has been determined in Ref \cite{Jenkins:2013zja,Jenkins:2013wua,Alonso:2013hga,Alonso:2014zka}. Considering the chosen relations in Eqn \ref{unconstrained}, we find the leading scale dependence
\bea
\mu \, \frac{d}{d \, \mu} (C_{HD} - 2  \, C^{(3)}_{\substack{Hl}}) = \frac{12 \, \lambda}{16 \, \pi^2} \, C_{HD} + \cdots
\eea
where  $\lambda = M_h^2/2 \, v^2$. Here we have assumed $C_{HD} - 2  \, C^{(3)}_{\substack{Hl}}$ vanishes at the scale $\mu \sim m_Z$, and we have neglected mixing
with other operators for simplicity. The dependence due to the top Yukawa accidentally cancels.
Numerically, running from the Z pole to $\sim 200 \, {\rm GeV}$ for LEP II $Z$ phenomenology, at least a percent level breaking of
this relation is already present. The leading breaking of the $C_{HD} -  C_{ll}$ chosen relation, neglecting mixing, is similarly 
\bea
\mu \, \frac{d}{d \, \mu} (C_{HD} - C_{ll}) = \frac{3}{4 \, \pi^2} \left(\lambda + y_t^2\right) \, C_{HD} + \cdots
\eea
There is some value in performing global EWPD fits, and not neglecting the scale dependence of the Wilson coefficients when considering flat directions. 
Phenomenology involving $V$ bosons at LHC is also not identical to $V$ phenomenology at LEP in this manner.

\section{Constructed Observables and basis choice}\label{pseudoobservablesection}

To further constrain the SMEFT, one can consider bounds on constructed observables.\footnote{A somewhat similar concept to constructed observables was recently discussed in Ref \cite{Gupta:2014rxa}.} The challenges of constructed  observables are well illustrated by the familiar oblique parameters, initially developed in Refs \cite{Kennedy:1988sn,Peskin:1990zt,Holdom:1990tc,Golden:1990ig,Altarelli:1990zd,Peskin:1991sw}. Using an oblique constraint as well as 
Eqn \ref{LEPdata1}-\ref{LEPdata4}  would be redundant. We first discuss oblique corrections as they illustrate the challenge
of constructed observables more directly than TGC verticies. 

In both cases, these
quantities are constructed with the assumption that the direct coupling of the $V$ to leptons is SM-like. 
Consider the consequences of this defining assumption for the effective axial and vector couplings in the SMEFT. 
With the normalization
\bea
\mathcal{L}_{chir}  =  (\bar{g}_1^2 + \bar{g}_2^2)^{1/2} J_\mu^0 Z^\mu + \frac{\bar{g}_2}{\sqrt{2} }J_\mu^\pm W_{\pm}^\mu,
\eea 
where $J_\mu^0 = \bar{\ell}_p \, \gamma_\mu \left(\bar{g}_V^{pr}- \bar{g}_A^{pr} \, \gamma_5 \right) \ell_r$, the shift in 
$\bar{g}_{V,A}$ in the standard basis (for charged leptons) are 
\bea\label{higherdgvga}
\bar{g}_V^{pr}&=& \left(\frac{1}{4} -  \bar{s}^2_\theta \right) \delta{g_V^{pr}}, \nn \\
\delta{g_V^{pr}} &=& 1 + \frac{v_T^2}{4 \, g_V^{SM}} \left(C_{HWB} \, \bar{s}_\theta \, \bar{c}_\theta \left(1-  4 \bar{c}_\theta^2 \right) + C_{\substack{H e \\ pr}} + C_{\substack{H \ell \\ pr}}^{(1)} + C_{\substack{H \ell \\ pr}}^{(3)}\right), \nn \\
\bar{g}_A^{pr}&=& \frac{1}{4} \delta{g_A^{pr}}, \nn \\
\delta{g_A^{pr}} &=& 1 + \frac{v_T^2}{4 \, g_A^{SM}} \left( C_{HWB} \, \bar{s}_\theta \, \bar{c}_\theta +  C_{\substack{H e\\ pr}}  
+ \left(C_{\substack{H \ell \\pr}}^{(1)} + C_{\substack{H \ell \\ pr}}^{(3)}\right) \frac{\left(1 + 2 \,\bar{s}_\theta^2 \right)}{1 - 2 \,\bar{s}_\theta^2} \right). 
\eea
Here $p.r$ are flavour labels. 
If the direct coupling of the $Z$ to leptons is SM-like, this naively corresponds to assuming
\bea\label{standardbasis-vanish}
C_{HWB} \, \bar{s}_\theta \, \bar{c}_\theta \left(1-  4 \bar{c}_\theta^2 \right) + C_{\substack{H e \\ pr}} + C_{\substack{H \ell \\ pr}}^{(1)} + C_{\substack{H \ell \\ pr}}^{(3)} \rightarrow 0, 
\nn \\ 
C_{HWB} \, \bar{s}_\theta \, \bar{c}_\theta +  C_{\substack{H e\\ pr}}  
+\frac{\left(1 + 2 \,\bar{s}_\theta^2 \right)}{1 - 2 \,\bar{s}_\theta^2}  \left(C_{\substack{H \ell \\pr}}^{(1)} + C_{\substack{H \ell \\ pr}}^{(3)}\right) \rightarrow 0,
\eea
while in the $\mathcal{O}$ basis this corresponds to assuming
\bea\label{obasis-vanish}
\frac{\left(\mathcal{P}_{B} + \mathcal{P}_{W}\right) \, g_1 \, g_2}{4} \, \bar{s}_\theta \, \bar{c}_\theta \left(1-  4 \bar{c}_\theta^2 \right) + C_{H e} 
+ \left[C_{\substack{H \ell \\ pr}}^{(1)} + C_{\substack{H \ell \\ pr}}^{(3)}\right]_{p \neq r} 
\rightarrow 0, 
\nn \\ 
\frac{\left(\mathcal{P}_{B} + \mathcal{P}_{W}\right) \, g_1 \, g_2}{4} \, \bar{s}_\theta \, \bar{c}_\theta +  C_{H e}
+  \frac{\left(1 + 2 \,\bar{s}_\theta^2 \right)}{1 - 2 \,\bar{s}_\theta^2} \left[C_{\substack{H \ell \\pr}}^{(1)} + C_{\substack{H \ell \\ pr}}^{(3)}\right]_{p \neq r}  
  \rightarrow 0.
\eea

The resulting constraints on the field theory when bounds on the oblique parameters are incorporated -- derived from experiments -- can be basis dependent
if this assumption is not imposed in a basis independent manner.\footnote{Notice that the number of constraints that the assumption corresponds to
is not even consistent between the bases, when flavour indicies are not ignored.} Expressing an observable
in terms of other observables is basis independent. Naively relating an observable to constructed observables is not.

\subsection{The $S$ parameter}\label{Sparameter}

In the PDG \cite{Beringer:1900zz1} (Sec.10) the $S$ parameter is defined as 
\bea\label{Sdefn}
\frac{\hat{\alpha}(m_Z)}{4 \, \hat{s}_Z^2 \, \hat{c}_Z^2} \, S \equiv 
\frac{\Pi_{ZZ}^{new}(m_Z^2) - \Pi_{ZZ}^{new}(0)}{m_Z^2}
 - \frac{\hat{c}_Z^2-\hat{s}_Z^2}{\hat{c}_Z \, \hat{s}_Z} \frac{\Pi^{new}_{Z \, \gamma}(m_Z^2)}{m_Z^2} -  \frac{\Pi^{new}_{\gamma \, \gamma}(m_Z^2)}{m_Z^2}.
\eea
The hatted parameters in Eqn \ref{Sdefn} are defined in the $\overline{\rm MS}$ scheme and conform with the PDG convention.
Assuming new physics is heavy enough for an operator interpretation, the $S$ parameter can be mapped to the Wilson coefficient of the operator $Q_{WB}$ \cite{Grinstein:1991cd}, as 
\bea
S_{\mathcal{Q}} = - \frac{16 \, \pi \, v_T^2}{g_1 \, g_2} \, C_{HWB}.
\eea
 In the $\mathcal{O}$ basis, the kinetic mixing of the 
photon and $Z$ due to higher dimensional operators is proportional to $\mathcal{P}_{B} + \mathcal{P}_{W}$; one finds
\bea\label{PbasisS}
S_{\mathcal{O}} = - 4 \, \pi \, v_T^2 \, \left(\mathcal{P}_{B} + \mathcal{P}_{W}\right).
\eea
Using the EoM relations between the Wilson coefficients to change basis
\bea\label{grojeansconfusion}
- 4 \, \pi \, v_T^2 \, \left(\mathcal{P}_{B} + \mathcal{P}_{W}\right) \rightarrow - \frac{16 \, \pi \, v_T^2}{g_1 \, g_2} \, C_{HWB}  - \frac{8 \, \pi \, v_T^2}{g_1^2 \, \hyp_\ell} \, C_{\substack{H \ell \\ tt}}^{(1)}  - \frac{16 \, \pi \, v_T^2}{g_2^2 \, \hyp_\ell} \, C_{\substack{H \ell \\ tt}}^{(3)}.
\eea
The idea of oblique parameters has an implicit challenge from field redefinitions, which is illustrated
by the above equation.
This is a point previously discussed, in part, in Ref \cite{SanchezColon:1998xg,Kilian:2003xt,Grojean:2006nn}.
At this stage it is important to note that even though the $S$ parameter in the two bases are related as in Eqn \ref{grojeansconfusion},
it {\it does not} directly follow that the $S$ parameter always has pure flat directions related to the operators $C_{H\ell}^{(1)},C_{H\ell}^{(3)}$ in the standard basis.
As explicitly demonstrated in Section \ref{constraints} the pure flat directions need not
be related to $C_{HWB}$.
 
Nevertheless, the definition of an oblique correction does have the defining assumption of a SM like $V$ coupling to leptons associated with it.\footnote{"Oblique" parameters are discussed in
Ref  \cite{Peskin:1991sw} as "Vaccum polarizations affect the above interactions by modifying the gauge-boson propagators.... 
This is the reason why they are called "oblique" corrections as opposed to the
"direct" vertex and box corrections that modify the form of the interactions themselves". } In both bases the operator $Q_{He}$ is present, so some version of this
defining assumption must always be imposed. 
One can consider a weak version of this defining condition, where only the combination of parameters present
in Eqn \ref{standardbasis-vanish} or Eqn \ref{obasis-vanish} are assumed to vanish.\footnote{This limit is fine tuned and is not invariant under the renormalization scale evolution of the theory.} 
A strong
version of an oblique parameter defining condition is to impose that each of the Wilson coefficients in Eqn \ref{higherdgvga} (other than the $C_{HWB}$) individually vanish. For the standard basis
this implies
\bea
C_{\substack{H e \\ pr}}, C_{\substack{H \ell \\ pr}}^{(1)}, C_{\substack{H \ell \\ pr}}^{(3)} \rightarrow 0.
\eea
 
The strong version of the oblique parameter defining assumption leads to the definitions in the bases being {\it identified}
\bea
-4 \, \pi \, v_T^2 \, \left(\mathcal{P}_{B} + \mathcal{P}_{W}\right) \equiv  - \frac{16 \, \pi \, v_T^2}{g_1 \, g_2} \, C_{HWB}.
\eea
So long as the standard definition of the oblique parameters \cite{Peskin:1991sw} is adhered to with the strong defining condition,
there is no issue with basis dependence. The weak version of this assumption results in the definitions still differing between bases.
This supports imposing the strong defining condition.
The PDG Higgs review \cite{Beringer:1900zz2} 
currently defines the oblique parameter $\Delta S$ in a basis dependent manner, proportional to $\mathcal{P}_{B} + \mathcal{P}_{W}$. This
is equivalent to the definition in Eqn \ref{Sdefn} in the PDG EW review \cite{Beringer:1900zz1} only when the strong version of the defining condition is imposed. 
Not imposing this condition changes the definition of this oblique parameter from its standard definition \cite{Kennedy:1988sn,Peskin:1990zt,Holdom:1990tc,Golden:1990ig,Altarelli:1990zd,Peskin:1991sw}, and introduces a basis dependent constructed observable. Using such a definition is inconsistent with basis independent bounds being obtained on the SMEFT.

Finally, we note that in the $\mathcal{O}$ basis, the strong LEP bound limit (including a strong constraint on the $S$ parameter) seems to correspond to $\mathcal{P}_{B} = - \mathcal{P}_{W}$ and $C_{H e} \rightarrow 0$. But this is an incomplete and basis dependent conclusion. Taking into account the EoM, and the strong LEP bound defining condition of the
$S$ parameter, $C_{\substack{H \ell \\ tt}}^{(1)},C_{\substack{H \ell \\ tt}}^{(3)} \rightarrow 0$ gives the non intuitive relationships 
$- \mathcal{P}_{HW} = \mathcal{P}_{HB} = \mathcal{P}_W  = - \mathcal{P}_B$ in the $\mathcal{O}$ basis. Not imposing this relationship
while assuming the strong LEP bound would lead to a functionally redundant operator basis.

\subsection{Triple Gauge coupling verticies}
Off shell TGC verticies are also not directly observable, like the oblique parameters, they are constructed observables.
The TGC vertex $Z \, W^+ \, W^-$ requires one of the massive gauge bosons to be off shell.
Leading experimental studies of this vertex result from measurements of 
\bea\label{TGCfinal}
\ell^+ \, \ell^- \rightarrow W^+ \, W^- \rightarrow jj \ell \, \nu, jjjj, jjX, \ell X,
\eea
 where $j$, $\ell$ and $X$ are a jet, lepton and missing final state energy \cite{Abdallah:2010zj,ALEPH:2005ab}. 
There are many ways to appreciate the distinction between the resulting constructed observable and the cross section measurement.
The kinematics of t and s channel exchange are distinct in $\sigma(e^+ \, e^- \rightarrow W^+ \, W^-)$.
The t-channel contribution dominates at threshold, however at high energies, the s-channel contribution related to the TGC vertex dominates \cite{Hagiwara:1986vm,Hagiwara:1992eh}.
The potential strength of TGC vertex bounds are directly related to the anomalous growth at high energies that results when the deviations from the 
SM in the s-channel are introduced. Using a reported bound for a TGC vertex for $Z \, W^+ \, W^-$,
the possible effect of $\mathcal{L}_6$ on the $t$-channel $\ell^+ \, \ell^- \rightarrow W^+ \, W^- $ process is set to zero.
To obtain the numerical values for the TGC bounds  \cite{Abdallah:2010zj,ALEPH:2005ab}, exclusive processes in Eqn \ref{TGCfinal} are assumed to have a SM like coupling of the $V$, and final states (including non-leptonic decays of the $W$) are combined, to improve statistics. This combination sets to zero possible modifications due to $\lsix$ in the decay channels. TGC verticies are clearly reported under the assumption that the possible effects of $\mathcal{L}_6$ on the direct coupling of the $V$ to leptons are set to zero.
It is important to reiterate that setting these contributions due to $\lsix$ to zero in constructing the observable is not equivalent to only removing the 
parameters that lead to these effects by field redefinitions. The defining condition must be mapped to the field theory using the EoM.

\subsubsection{TGC results}
TGC verticies have recently come under renewed scrutiny for the SMEFT in Refs.~\cite{Eboli:2010qd,Corbett:2012ja,Corbett:2013pja,Buchalla:2013wpa}.
These analyses descend from the classic works on higher dimensional operators in TGC's \cite{Hagiwara:1986vm,DeRujula:1991se,Hagiwara:1992eh}
that introduced the standard notation \cite{Hagiwara:1986vm}
\bea
(-\mathcal{L}_{TGC}) &=& i \, \bar{g}_2 \left[\delta g_1^Z \, \bar{c}_\theta \, \mathcal{Z}^\mu \left(\mathcal{W}^{+}_{\mu \nu} \, \mathcal{W}^{-\nu} - \mathcal{W}^{-}_{\mu \nu} \, \mathcal{W}^{+\nu}\right) \right]
+ i\, \bar{g}_2 \left[\delta g_1^\gamma \, \bar{s}_\theta \, \mathcal{A}^\mu \left(\mathcal{W}^{+}_{\mu \nu} \, \mathcal{W}^{-\nu} - \mathcal{W}^{-}_{\mu \nu} \, \mathcal{W}^{+\nu}\right) \right], \nn \\
&+& i \, \bar{g}_2 \left[\delta \kappa_Z \, \bar{c}_\theta \, \mathcal{Z}_{\mu \nu} \, \mathcal{W}^{+\nu} \, \mathcal{W}^{-\mu} + \delta \kappa_\gamma \, \bar{s}_\theta \mathcal{A}_{\mu \nu} \mathcal{W}^{+\nu} \, \mathcal{W}^{-\mu} \right], \nn \\
&+& i \, \frac{\bar{g}_2}{\bar{m}_W^2} \, \left(\lambda_Z \, \bar{c}_\theta \mathcal{Z}^{\mu \, \nu} +\lambda_\gamma \, \bar{s}_\theta \mathcal{A}^{\mu \, \nu} \right) \left(W^+_{\rho \, \mu} W^{- \rho}_{\nu} \right).
\eea
Here the field strengths for the massive gauge bosons are using the short hand notation $V_{\mu \, \nu} = \partial_\mu V_\nu -\partial_\nu V_\mu$,
and a number of other notational conventions are present. The mass eigenstate gauge bosons in $\mathcal{L}_{SM} + \lsix$ are denoted $\mathcal{Z},\mathcal{A},\mathcal{W}$. (See Ref \cite{Alonso:2013hga} Section 5.4 for the explicit definitions.)
Note that the lagrangian parameters in the canonically normalized SMEFT, $\bar{g}_2$, $\bar{c}_\theta$, $\bar{s}_\theta$ are present defining the anomalous parameters
$\delta g_1^{Z, \gamma}, \delta \kappa_{Z, \gamma}, \lambda_{Z, \gamma}$. The overall sign convention is consistent with Ref \cite{Hagiwara:1986vm}, indicated in the above equation with an explicit $-\mathcal{L}_{TGC}$, which is opposite the overall sign convention in Refs. \cite{Jenkins:2013zja,Alonso:2013hga}. 
The $\mathcal{L}_{SM}+ \lsix$ TGC anomalous couplings in the standard basis are given by 
\begin{align}\label{standardTGC2}
\delta g_1^{Z} &=  \frac{\bar{s}_\theta \, v_T^2}{2 \, \bar{c}_\theta}  \, C_{HWB},      & 
\delta g_1^{\gamma} &= - \frac{\bar{c}_\theta \, v_T^2}{2 \, \bar{s}_\theta} \,  C_{HWB},    \\
\delta \kappa_{Z} &= - \frac{\bar{s}_\theta \, v_T^2}{2 \, \bar{c}_\theta}  \, C_{HWB},     &
\delta \kappa_{\gamma} &= \frac{\bar{c}_\theta \, v_T^2}{2 \, \bar{s}_\theta} \,  C_{HWB},     \\
\delta \lambda_{Z} &=6 \, \frac{\bar{m}_W^2}{\Lambda^2}  \, C_{W}, &\delta \lambda_{\gamma} &=6 \, \frac{\bar{m}_W^2}{\Lambda^2}  \, C_{W}.  
\end{align}
Note that these results are expressed in terms of the canonically normalized Lagrangian parameters, including $\bar{c}_\theta,\bar{s}_\theta$ as defined in Ref \cite{Alonso:2013hga}. A redefinition of the effective mixing angle, to absorb a shift due to $C_{HWB}$, has not yet been done. 
The $\mathcal{L}_{SM}+ \lsix$ TGC anomalous couplings in the $\mathcal{O}$ basis are given by \cite{Pomarol:2013zra} 
\begin{align}\label{PRresult}
\delta g_1^{Z} &= \frac{v_T^2 \,  \bar{g}_1^2 \, \left(\mathcal{P}_{B} + \mathcal{P}_{W}\right)}{8} + \frac{(\bar{g}_1^2 + \bar{g}_2^2) \, v_T^2}{4} \left(\mathcal{P}_{HW} + \mathcal{P}_{W}\right),  \\
\delta g_1^{\gamma} &= - \frac{v_T^2 \,  \bar{g}_2^2 \, \left(\mathcal{P}_{B} + \mathcal{P}_{W}\right)}{8},    \\
\delta \kappa_{Z} &= \frac{v_T^2 \,  \bar{g}_1^2 \, \left(\mathcal{P}_{B} + \mathcal{P}_{W}\right)}{8} + \frac{v_T^2}{4 } \, (\bar{g}_1^2 + \bar{g}_2^2) \, \left(\mathcal{P}_{HW} + \mathcal{P}_{W}\right) - \frac{v_T^2}{4 } \, g_1^2 \left(\mathcal{P}_{HW} + \mathcal{P}_{HB}\right),   \\
\delta \kappa_{\gamma} &= - \frac{v_T^2 \,  \bar{g}_2^2 \, \left(\mathcal{P}_{B} + \mathcal{P}_{W}\right)}{8} + \frac{g_2^2 \, v_T^2}{4} \, \left(\mathcal{P}_{HB} + \mathcal{P}_{HW}\right),     \\
\delta \lambda_{Z} &=6 \, \frac{\bar{m}_W^2}{\Lambda^2}  \, C_{W}, \\
 \delta \lambda_{\gamma}&=6 \, \frac{\bar{m}_W^2}{\Lambda^2}  \, C_{W}.
\end{align}
The operator $Q_W = \epsilon_{IJK} \, W_\mu^{I,\nu}  \, W_\nu^{J,\rho}  \, W_\rho^{K,\mu}$ 
is not to be confused with the operator $\mathcal{O}_W$ in the $\mathcal{O}$ basis. Including this operator,
leads to a flat direction in constraints derived from TGC verticies \cite{Brooijmans:2014eja,Ellis:2014dva}, as expected \cite{Alonso:2013hga}.\footnote{The PDG Higgs review \cite{Beringer:1900zz2}  treatment of this flat direction is unspecified, and this operator is not included in the quoted TGC bounds in the PDG.} The 
mixing angles have not been related to input observables as yet in Eqn \ref{PRresult}. Doing so the dependence on $\mathcal{P}_{B} + \mathcal{P}_{W}$ is removed and the
expressions satisfy $\delta \kappa_Z = \delta g^Z_1 - t_{\bar{\theta}}^2 \, \delta \kappa_\gamma$ in both bases, as expected \cite{Hagiwara:1986vm}.

\subsubsection{Relation to input observables}\label{TGCredefine}
One can absorb the redefinition of the mixing angles in the SMEFT in a finite renormalization. This takes into account how the dependence on
$C_{HWB}$ modifying the mixing angle cancels when relating TGC verticies to input observables.
Doing so, the deviations in $\delta g_1^Z, \delta g_1^{\gamma} $ due to $C_{HWB}$ are canceled, and 
\begin{align}
\delta \tilde{\kappa}_{Z} &=  1 + \frac{\bar{s}_\theta \, v_T^2}{\bar{c}_\theta}  \, C_{HWB}, & \delta \tilde{\kappa}_{\gamma} &=1 - \frac{\bar{c}_\theta \, v_T^2}{\bar{s}_\theta} \,  C_{HWB}.
\end{align}
In Eqn. \ref{standardTGC2}, the Wilson coefficients $C_{HWB}$ and $C_W$ are present. The Wilson coefficient $C_{HWB}$ need not be related to a pure flat direction in the standard basis. Exchanging $\bar{g}_2$ in terms of $m_W$,
introduces the parameter shifts $\delta m_W$ and $\delta G_F$. However, the former is already used as a measurement in Eqn \ref{LEPdata4} and the flat direction can be chosen
to set $\delta G_F =0$, as demonstrated. Similarly
exchanging the mixing angles in terms of input parameters cancels the deviations in $\delta g_1^Z, \delta g_1^{\gamma}$  but does not introduce sensitivity
to the remaining flat directions.

The defining conditions of the off-shell TGC bounds are inconsistent with choices that can be made for flat directions present due to 
LEP data. These directions can be chosen so that it is crucial to probe $C_{H\ell}^{(3)}$ to break the remaining degeneracy, see Eqn \ref{unconstrained}.
Breaking this degeneracy can be done by studying {\it exclusive} $W$ decay to leptonic final states, as 
$ \delta \Gamma_W/\Gamma_W \propto \bar{g}_2 \, v_T^2 C_{\substack {H\ell \\ tt}}^{(3)}$.  For example, a process that can
remove a degeneracy is exclusive $\sigma(e^+ \, e^- \rightarrow \ell \, \bar{\ell} X)$.
Inclusive $\sigma(e^+ \, e^- \rightarrow W^+ \, W^-)$ production that includes the $\nu$ t-channel exchange can also be used. 
Bounds on the off-shell TGC vertex do not directly probe these effects, and their defining assumptions assume these effects
in the SMEFT are set to zero.

The conclusion that TGC verticies are limited in their utility holds in the $\mathcal{O}$ basis, but the reasoning is more subtle and involves a functional redundancy. 
Examining the EoM relations one finds
\begin{align}
\mathcal{P}_{HW}+\mathcal{P}_{HB} &\rightarrow - \frac{4}{\bar{g}_1 \, \bar{g}_2} C_{HWB}, & \mathcal{P}_{HW}+\mathcal{P}_{W} &\rightarrow \frac{4}{\bar{g}_2^2} C_{\substack{H\ell \\ tt}}^{(3)}.
\end{align}

Using TGC constructed observables to bound a parameter equivalent to $C_{H\ell}^{(3)}$ is functionally redundant.
Analyses that use these  constructed observables can constrain the field theory in a  consistent manner, when the defining assumptions of the TGC verticies are imposed in a basis independent manner, avoiding a functional redundancy. In this case $ \mathcal{P}_{HW}+\mathcal{P}_{W}  \rightarrow 0$.\footnote{One can always choose that the flat directions do correspond to the TGC constructed observables, and the $\mathcal{O}$ basis makes such a choice intuitive. However, this does not establish
the general utility of this constructed observable, but reinforces the inconsistency of incorrectly treating it as a measurement. How strongly constrained an EFT is cannot depend on an arbitrary operator basis, or flat direction, choice.} Measurements of $\sigma(e^+ \,e^- \rightarrow W^+ \, W^-)$,
are sensitive at the $\sim 1 \%$ level to deviations in the coupling of the $W$, so no pure flat directions are expected in a full analysis 
using the observables that can lift the flat direction consistently. The nature of the exact numerical bound is worthy of future 
study.\footnote{It is interesting to note that allowed deviations in the $h \rightarrow V \, F$ spectra are de-correlated in 
the case of the nonlinear EFT from LEP measurements. This makes accurate and precise studies of the $h \rightarrow V \, F$ spectra, in light of consistent LEP constraints, particularly interesting \cite{Isidori:2013cga,Brivio:2013pma}.}


\section{Triple Gauge coupling verticies and $h \rightarrow V \mathcal{F}$}\label{hvfspec}

In this section, we reexamine the relationship between reported bounds on TCG verticies and the $h \rightarrow V \, F$ differential distributions. 
We ensure that the defining condition of the  TGC constructed observable is also imposed consistently when considering this relationship by adopting
the strong LEP limit. We demonstrate how accounting for the the subtlety of the functional redundancy, and considering the EoM
makes the connection between these observables vanish in the limit of strong LEP bounds.
The importance of the $h \rightarrow V \, F$ differential distributions has recently been studied in Refs.~\cite{DeRujula:2010ys,Gainer:2013rxa,Isidori:2013cla,Grinstein:2013vsa,Buchalla:2013mpa,Beneke:2014sba,Gonzalez-Alonso:2014rla}.
The relationship between these quantities has received some attention in Refs \cite{Pomarol:2013zra,Brooijmans:2014eja,Ellis:2014dva}. The arguments of 
Ref.\cite{Pomarol:2013zra} have been influential and have lead to claims in the recent Higgs review of the 
PDG \cite{Beringer:1900zz2}.

We focus on the case when $V = Z$, although the same arguments apply for $V = W$. 
In the SM, the result for the offshell gauge boson invariant mass ($q^2$) distribution is given by 
\bea
\frac{d\Gamma_0(\hat{q}^2)}{d \hat{q}^2} =  \frac{ (\bar{g}_1^2 + \bar{g}_2^2)^2 \, (g_A^2 + g_V^2) \, m_h \,\left[\lambda^2(\hat{q}^2,\rho) + 12 \rho \, \hat{q}^2 \right] }{256 \pi^3} 
\frac{\lambda(\hat{q}^2,\rho)}{\left(\hat{q}^2 - \rho \right)^2}. 
\eea
Here $\hat{q}^2 = q^2/m_h^2$ and $\rho = m_V^2/m_h^2$. The masses here are the physical (measurable) 
on shell masses of the vector bosons and $\lambda(\hat{q}^2,\rho) = \sqrt{(1+ \hat{q}^2 - \rho)^2 - 4 \hat{q}^2}$. The modification of the $q^2$ distribution 
due to $\lsix$ is given by
\bea\label{Zspectra2}
\frac{1}{v_T^2}\frac{ \rd \Gamma }{\rd  q^2 } &=& \frac{1}{3} \frac{ \rd \Gamma_0 }{\rd  \hat{q}^2 }   \, \biggl\{\frac{1}{v_T^2} + 2 C_{H \Box} + C_{HD}  + \frac{g_V + g_A}{(g_V)^2 + (g_A)^2} \frac{C_{He}}{2} + \frac{2 \, g_1 \, g_2}{g_1^2 + g_2^2} C_{HWB} \biggl\} ,  \nn \\
&+& \frac{1}{3} \frac{ \rd \Gamma_0 }{\rd  \hat{q}^2 }   \, \biggl\{ \frac{\bar{s}_\theta  \, \bar{c}_\theta C_{HWB} }{2 \, (g_A^2 + g_V^2)} \left(g_A + g_V (1 - 4 \, \bar{c}_\theta^2) \right) +  \frac{C_{H \ell}^{(1)}+ C_{H \ell}^{(3)}}{2 \, \left(g_V^2 + g_A^2 \right)} \left(g_V - g_A \, \frac{2 \, \bar{s}_\theta^2 + 1}{2 \, \bar{s}^2_\theta - 1} \right)   \biggl\},  \nn \\
&+& 8 \, \hat{q}^2 \,  \frac{ \rd \Gamma_0}{\rd  \hat{q}^2 } \, \left[ \bar{s}_\theta  \, \bar{c}_\theta\, C_{HWB}  \, + C_{HB} \, \bar{s}^2_\theta + C_{HW} \, \bar{c}^2_\theta \right] \,  \left( \frac{\hat{q}^2 -1 + \rho}{\lambda^2(\hat{q}^2,\rho) + 12 \rho \, \hat{q}^2} \right), \nn \\ 
&+& \frac{256 \, \pi \, \alpha_{ew} \, \bar{s}_\theta \, \bar{c}_\theta}{\bar{g}_1^2 + \bar{g}_2^2} \,  \frac{ \rd \Gamma_0}{\rd  \hat{q}^2 } \, \mathscr{C}_{\gamma Z} \,\left( \frac{g_V}{(g_V)^2 + (g_A)^2}\right) \, 
\left(\frac{\rho \, (\rho - \hat{q}^2) \, (\hat{q}^2 -1 + \rho)}{\lambda^2(\hat{q}^2,\rho) + 12 \rho \, \hat{q}^2} \right), \nn \\ 
&-& \frac{ \rd \Gamma_0}{\rd  \hat{q}^2 } \, \frac{(\rho
 - \hat{q}^2)}{6 \, \rho} \, \left(C_{He} \frac{\left(g_V - g_A \right)}{\left(g_V^2 + g_A^2 \right)} + (C_{H \ell}^{(1)}+ C_{H \ell}^{(3)})\frac{\left(g_V + g_A \right)}{\left(g_V^2 + g_A^2 \right)}  \right).
\eea
We have explicitly labelled the term that comes from the photon pole exchange 
with $\mathscr{C}_{\gamma Z}$.The
Wilson coefficients for $h \rightarrow \gamma \, Z, h \rightarrow \gamma \, \gamma$ are defined with the normalization in Ref \cite{Alonso:2013hga}.
A consistent scheme can include the squared photon pole contribution \cite{Beneke:2014sba}, however, for the sake of our illustrative discussion on the EoM effects, we neglect this term.  

In the case of strong experimental LEP bounds, it has been argued that the $h \rightarrow V \, F$ offshell invariant mass ($q^2$) spectrum is not 
a competitive source of information on higher dimensional operators due to their relationship with TGC verticies. In this limit, 
\bea\label{Zspectra6}
\frac{1}{v_T^2}\frac{ \rd \Gamma }{\rd  q^2 } &=&  \frac{ \rd \Gamma_0 }{\rd  \hat{q}^2 }  \, \mathcal{N} \left(0, C_{H \Box}, C_{HD},0,0, 0  \right),  \nn \\
&+& 32 \, \pi \, \alpha_{ew} \,  \frac{ \rd \Gamma_0}{\rd  \hat{q}^2 } \,  \mathscr{C}_{\gamma \gamma} \,  \hat{q}^2  \, \left(\frac{\hat{q}^2 -1 + \rho}{\lambda^2(\hat{q}^2,\rho) + 12 \rho \, \hat{q}^2} \right), \nn \\ 
&+& 32 \, \pi \, \alpha_{ew} (\cot_{\bar{\theta}} - \tan_{\bar{\theta}})\frac{ \rd \Gamma_0}{\rd  \hat{q}^2 } \,\mathscr{C}_{\gamma Z} \, \hat{q}^2  \, \left(\frac{\hat{q}^2 -1 + \rho}{\lambda^2(\hat{q}^2,\rho) + 12 \rho \, \hat{q}^2} \right), \nn \\ 
&+& \frac{256 \, \pi \, \alpha_{ew} \, \bar{s}_\theta \, \bar{c}_\theta}{\bar{g}_1^2 + \bar{g}_2^2} \,  \frac{ \rd \Gamma_0}{\rd  \hat{q}^2 } \, \mathscr{C}_{\gamma Z} \,\left( \frac{g_V}{(g_V)^2 + (g_A)^2}\right) \, 
\left(\frac{\rho \, (\rho - \hat{q}^2) \, (\hat{q}^2 -1 + \rho)}{\lambda^2(\hat{q}^2,\rho) + 12 \rho \, \hat{q}^2} \right),
\eea
where a normalization function, 
$\mathcal{N}(C_{HWB}, C_{H \Box}, C_{HD},C_{He},C_{H \ell}^{(1)},C_{H \ell}^{(3)})$,
has been introduced. In the strong LEP limit, the BSM momentum dependence of this spectra is directly related to measurements of $\mathscr{C}_{\gamma \gamma},\mathscr{C}_{\gamma Z}$. However in this same limit, this spectrum is not related to TGC verticies. The functional form of the shape dependent deviation in the spectrum due to $C_{\gamma Z}$ is given in Eqn \ref{Zspectra6}, and can be fit for in dedicated searches.
Considering the relative experimental accessibility of $h \rightarrow \gamma Z$ and the $h \rightarrow V \, F$ spectra, the latter spectra can be thought of a leading indirect probe of $\mathscr{C}_{\gamma Z}$. 

In the $\mathcal{O}$ basis the spectrum of interest is given by 
\bea\label{ZspectraP}
\frac{1}{v_T^2}\frac{ \rd \Gamma }{\rd  q^2 } &=& \frac{1}{3} \frac{ \rd \Gamma_0 }{\rd  \hat{q}^2 }   \, \biggl\{\frac{1}{v_T^2} + 2 C_{H \Box} - \frac{\mathcal{P}_{T}}{4}  + \frac{g_V + g_A}{(g_V)^2 + (g_A)^2} \frac{C_{He}}{2} + \frac{g_1^2 \, g_2^2}{2 (g_1^2 + g_2^2)} (\mathcal{P}_B + \mathcal{P}_W)\biggl\} ,  \nn \\
&+&  \frac{1}{24} \frac{ \rd \Gamma_0 }{\rd  \hat{q}^2 } \left( \mathcal{P}_{B} + \mathcal{P}_{W} \right) \frac{\bar{s}^2  \, \bar{c}^2 (g_1^2 + g_2^2) \left[g_A + g_V(1 - 4 \, \bar{c}^2) \right] }{\left(g_V^2 + g_A^2 \right)} ,  \nn \\
&+& 8 \, \hat{q}^2 \,  \frac{ \rd \Gamma_0}{\rd  \hat{q}^2 } \, \left[ \bar{s}^2 \, C_{HB}  \, -  \frac{(\mathcal{P}_{HB} \, g_1^2 + \mathcal{P}_{HW} \, g_2^2)}{4} 
 \frac{\left(g_A^2 - g_V^2\right)}{\left(g_V^2 + g_A^2 \right)} \right] \,  \left(\frac{\hat{q}^2 -1 + \rho}{\lambda^2(\hat{q}^2,\rho) + 12 \rho \, \hat{q}^2} \right), \nn \\ 
&-& 4 \, \frac{\bar{s}^2_\theta \, \bar{c}^2_\theta \, m_h^2}{v_T^2}  \frac{ \rd \Gamma_0}{\rd  \hat{q}^2 } \, (\mathcal{P}_{HB} - \mathcal{P}_{HW}) \,\left( \frac{g_V}{(g_V)^2 + (g_A)^2}\right) \, 
\left(\frac{\rho \, (\rho - \hat{q}^2) (\rho + \hat{q}^2 -1)}{\lambda^2(\hat{q}^2,\rho) + 12 \rho \, \hat{q}^2} \right), \nn \\ 
&-& \frac{2 \, \bar{s}^2  \, \bar{c}^2 \, m_h^2}{3 \, v_T^2} \frac{ \rd \Gamma_0}{\rd  \hat{q}^2 } \, (\rho - \hat{q}^2) \left( \frac{g_V}{(g_V)^2 + (g_A)^2}\right) \left( \mathcal{P}_{HB} +\mathcal{P}_{B} -\mathcal{P}_{HW} -\mathcal{P}_{W} \right), \nn \\
&+& \frac{2 \, m_h^2}{3 \, v_T^2} \frac{ \rd \Gamma_0}{\rd  \hat{q}^2 } \, (\rho + \hat{q}^2) \left( \bar{s}^2 (\mathcal{P}_{HB} +\mathcal{P}_{B}) + \bar{c}^2 (\mathcal{P}_{HW} +\mathcal{P}_{W})\right), \nn \\
&-& \frac{ \rd \Gamma_0}{\rd  \hat{q}^2 } \, \frac{(\rho - \hat{q}^2)}{6 \, \rho} \, \left(C_{He} \frac{\left(g_V - g_A\right)}{\left(g_V^2 + g_A^2 \right)} \right).
\eea

Taking into account the EoM subtlety in the strong LEP limit, imposed to use constraints due to TGC vertex bounds, one finds
\begin{align}
\mathcal{P}_{B} + \mathcal{P}_{W} &\rightarrow 0, \\
\mathcal{P}_{HW} + \mathcal{P}_{W} &\rightarrow 0,  \\
\mathcal{P}_{HB} + \mathcal{P}_{B} &\rightarrow 0,  \\
\mathcal{P}_{HW} + \mathcal{P}_{HB} &\rightarrow 0, \\
\mathcal{P}_{HB} +\mathcal{P}_{B} -\mathcal{P}_{HW} -\mathcal{P}_{W} &\rightarrow 0.
\end{align}
Consistent between the bases, the TGC verticies are not related to $h \rightarrow V \, F$ measurements in this limit.
The combinations of Wilson coefficients that vanish in the strong LEP limit appear frequently in calculations using the $\mathcal{O}$ basis.

\section{Conclusions}\label{conclude}

There are 2499 free parameters in the dimension six operator corrections to the SM in the SMEFT. As such, it is inevitable that theoretical and experimental assumptions will be made to simplify the study of the SMEFT.
Although this can be done in a consistent manner using approximate symmetries that constrain the S matrix, it is likely that constructed
observables will also be used.

Any operator basis can be used to study the SMEFT and no basis is superior or inferior to any other. 
At the same time, it is an unfortunate fact that the potential for a functional redundancy in the $\mathcal{O}$ basis
is directly related to imposing the assumption of a SM like $V$ coupling to leptons in future experimental studies, i.e the limit of strong LEP constraints
in constructed LHC observables. 

We have illustrated the issues involved in avoiding the potential inconsistencies of constructed observables considering the oblique parameters,
TGC verticies, and the relation between the TGC verticies and the $h \rightarrow V \, F$ spectra. Using multiple bases, and keeping note of the EoM relations
between bases can make the non intuitive constraints, and defining conditions, of constructed observables transparent. 

As the data set from LHC advances, ever more complicated final states will be studied, and derived constraints -- or deviations -- in such measurements
will be incorporated into the SMEFT. It is essential that such studies are performed in a consistent and basis independent manner when constructed observables are used.

\section*{Acknowledgements}

We thank Martin Gonzalez-Alonso, Cliff Burgess and particularly Gino Isidori for insightful and useful discussions related to this work. We also thank  Rodrigo Alonso, Cliff Burgess, Ben Grinstein, Aneesh Manohar
and Gino Isidori for comments on the manuscript.

\appendix
\section{The SM Equation of Motion}

As a consequence of the principle of least action, the SM fields satisfy the classical EoM.
Field redefinitions can be used to eliminate redundant operators when changes in $\mathcal{L}$, due to a field redefinition, lead to shifts in $S$ matrix elements
that vanish by the EoM. For the Higgs field, the SM EoM is given by
\begin{align}
D^2 H_k -\lambda v^2 H_k +2 \lambda (H^\dagger H) H_k + \overline q^j\, Y_u^\dagger\, u \, \epsilon_{jk} + \overline d\, Y_d\, q_k +\overline e\, Y_e\,  l_k =0. 
\label{eomHiggs}
\end{align}
The derivatives acting on the SM fermion fields have the EoM's
\begin{align}
i\slashed{D}\, q_j &= Y_u^\dagger\, u\, \widetilde H_j + Y_d^\dagger\, d\, H_j \,, &
i\slashed{D}\,  d &= Y_d\,  q_j\, H^{\dagger\, j} \,, &
i\slashed{D}\, u &= Y_u\, q_j\, \widetilde H^{\dagger\, j}\,, \nn \\
i\slashed{D} \, l_j &= Y_e^\dagger\, e  H_j  \,, &
i\slashed{D} \, e &= Y_e\, l_j H^{\dagger\, j}.
\label{eompsifields}
\end{align}
Finally for the gauge fields, the EoM are 
\begin{align}
\frac{1}{g_3} \, \left[D^\alpha , G_{\alpha \beta} \right]^A &= \sum_{\psi=u,d,q} \overline \psi \, T^A \gamma_\beta  \psi,  \nn \\
\frac{1}{g_2} \, \left[D^\alpha , W_{\alpha \beta} \right]^I &=  \frac 12 \overline q \, \tau^I \gamma_\beta  q + \frac12 \overline l \, \tau^I \gamma_\beta  l + \frac12 H^\dagger \, i\overleftrightarrow D_\beta^I H\,, \nn \\
\frac{1}{g_1} \, D^\alpha B_{\alpha \beta} &= \sum_{\psi=u,d,q,e,l} \overline \psi \, \hyp_i \gamma_\beta  \psi + \frac12 H^\dagger \, i\overleftrightarrow D_\beta H.
\label{eomX}
\end{align}
Here $\left[D^\alpha , F_{\alpha \beta} \right]$ is the covariant derivative in the adjoint representation and $\hyp_i$ are the $U(1)$ hypercharges of the 
fermions.\footnote{We have used the derivative notation
\begin{align}
H^\dagger \, i\overleftrightarrow D_\beta H &= i H^\dagger (D_\beta H) - i (D_\beta H)^\dagger H \,,\nn \\
H^\dagger \, i\overleftrightarrow D_\beta^I H &= i H^\dagger \tau^I (D_\beta H) - i (D_\beta H)^\dagger \tau^I H.
\end{align}}
The EoM relate operators with a different set of fields and can lead to non intuitive physics, which is required for the basis independence of the field theory. 

\section{Notation}

The Lagrangian we use is given by $\mathcal{L}=\mathcal{L}_{\rm SM}+\lsix$. To establish notation, the SM Lagrangian\footnote{This appendix is largely already stated in Ref\cite{Jenkins:2013wua} and is restated here to make the arguments of the paper more self contained.} is given as
\begin{align}
\mathcal{L} _{\rm SM} &= -\frac14 G_{\mu \nu}^A G^{A\mu \nu}-\frac14 W_{\mu \nu}^I W^{I \mu \nu} -\frac14 B_{\mu \nu} B^{\mu \nu}
+ (D_\mu H^\dagger)(D^\mu H)
+\sum_{\psi=q,u,d,l,e} \overline \psi\, i \slashed{D} \, \psi\nn \\
&-\lambda \left(H^\dagger H -\frac12 v^2\right)^2- \biggl[ H^{\dagger j} \overline d\, Y_d\, q_{j} + \widetilde H^{\dagger j} \overline u\, Y_u\, q_{j} + H^{\dagger j} \overline e\, Y_e\,  l_{j} + \hbox{h.c.}\biggr].
\label{sm}
\end{align}
$H$ is an $SU(2)$ scalar doublet with hypercharge $\hyp_H=1/2$. The Higgs boson mass is given as $m_H^2=2\lambda v^2$, with $v \sim 246$~GeV. Fermion mass matrices are $M_{u,d,e}=Y_{u,d,e}\, v /\sqrt 2$. The covariant derivative is $D_\mu = \partial_\mu + i g_3 T^A A^A_\mu + i g_2  t^I W^I_\mu + i g_1 \hyp B_\mu$. Here $T^A$ are $SU(3)$ generators,  $t^I=\tau^I/2$ are $SU(2)$, and $\hyp$ is the $U(1)$ hypercharge generator.  $SU(2)$ indices have the convention $j,k$ and $I,J,K$ for the fundamental and adjoint, respectively. The $SU(3)$ indices $A,B,C$ are in the adjoint representation. $\widetilde H$ is defined by $H_j = \epsilon_{jk} H^{\dagger\, k}$
where the $SU(2)$ invariant tensor $\epsilon_{jk}$ is defined by $\epsilon_{12}=1$ and $\epsilon_{jk}=-\epsilon_{kj}$, $j,k=1,2$.  Fermion fields $q$ and $l$ are left-handed fields, and $u$, $d$ and $e$ are right-handed fields. We use $p,r,s,t$ for flavor indices (each of which run over the three generations) which are suppressed in Eq.~(\ref{sm}). 
The Yukawa matrices $Y_{u,d,e}$ are matrices in flavor space, as are some operator Wilson coefficients.
The flavour index convention used is explicitly given in Section 2.1 of Ref \cite{Jenkins:2013zja}.
The main notational change from Ref.~\cite{Grzadkowski:2010es} is the replacement of $\varphi$ by $H$ for the Higgs field. We use the convention $\widetilde F_{\mu \nu} =(1/2) \epsilon_{\mu \nu \alpha \beta} F^{\alpha \beta}$ with $\epsilon_{0123}=+1$. We relist the operators given in Ref.~\cite{Grzadkowski:2010es} here for completeness.

In using the SMEFT, we take the theory to canonical form, introducing "bar" labels onto the standard model parameters, such as $\bar{g}_{1,2}$. 
All of the steps to do this are discussed in Ref \cite{Alonso:2013hga} in Section 5.4. Some of the parameters in the SMEFT are explicitly defined as follows. 
The modified potential is
\begin{align}
V(H) &= \lambda \left(H^\dagger H -\frac12 v^2\right)^2 - C_H \left( H^\dagger H \right)^3,
\label{pot}
\end{align}
yielding the new minimum
\begin{align}
\langle H^\dagger H \rangle &= \frac{v^2}{2} \left( 1+ \frac{3 C_H v^2}{4 \lambda} \right) \equiv \frac12 v_T^2,
\end{align}
while the mixing angles are
\begin{align}
\sin \tc &= \frac{\gcb}{\sqrt{\gcb^2+\gcw^2}}\left[1 + \frac{v_T^2}{2}  \,   \, \frac{\gcw}{\gcb}\ \frac{\gcw^2-\gcb^2}{\gcw^2+\gcb^2} C_{HWB} \right], \nn \\
\cos \tc &= \frac{\gcw}{\sqrt{\gcb^2+\gcw^2}}\left[1 - \frac{v_T^2}{2}  \,   \, \frac{\gcb}{\gcw}\ \frac{\gcw^2-\gcb^2}{\gcw^2+\gcb^2} C_{HWB} \right].
\end{align}
%
\begin{table}
\begin{center}
\small
\begin{minipage}[t]{4.45cm}
\renewcommand{\arraystretch}{1.5}
\begin{tabular}[t]{c|c}
\multicolumn{2}{c}{$1:X^3$} \\
\hline
$Q_G$                & $f^{ABC} G_\mu^{A\nu} G_\nu^{B\rho} G_\rho^{C\mu} $ \\
$Q_{\widetilde G}$          & $f^{ABC} \widetilde G_\mu^{A\nu} G_\nu^{B\rho} G_\rho^{C\mu} $ \\
$Q_W$                & $\epsilon^{IJK} W_\mu^{I\nu} W_\nu^{J\rho} W_\rho^{K\mu}$ \\ 
$Q_{\widetilde W}$          & $\epsilon^{IJK} \widetilde W_\mu^{I\nu} W_\nu^{J\rho} W_\rho^{K\mu}$ \\
\end{tabular}
\end{minipage}
\begin{minipage}[t]{2.7cm}
\renewcommand{\arraystretch}{1.5}
\begin{tabular}[t]{c|c}
\multicolumn{2}{c}{$2:H^6$} \\
\hline
$Q_H$       & $(H^\dag H)^3$ 
\end{tabular}
\end{minipage}
\begin{minipage}[t]{5.1cm}
\renewcommand{\arraystretch}{1.5}
\begin{tabular}[t]{c|c}
\multicolumn{2}{c}{$3:H^4 D^2$} \\
\hline
$Q_{H\Box}$ & $(H^\dag H)\Box(H^\dag H)$ \\
$Q_{H D}$   & $\ \left(H^\dag D_\mu H\right)^* \left(H^\dag D_\mu H\right)$ 
\end{tabular}
\end{minipage}
\begin{minipage}[t]{2.7cm}

\renewcommand{\arraystretch}{1.5}
\begin{tabular}[t]{c|c}
\multicolumn{2}{c}{$5: \psi^2H^3 + \hbox{h.c.}$} \\
\hline
$Q_{eH}$           & $(H^\dag H)(\bar l_p e_r H)$ \\
$Q_{uH}$          & $(H^\dag H)(\bar q_p u_r \widetilde H )$ \\
$Q_{dH}$           & $(H^\dag H)(\bar q_p d_r H)$\\
\end{tabular}
\end{minipage}

\vspace{0.25cm}

\begin{minipage}[t]{4.7cm}
\renewcommand{\arraystretch}{1.5}
\begin{tabular}[t]{c|c}
\multicolumn{2}{c}{$4:X^2H^2$} \\
\hline
$Q_{H G}$     & $H^\dag H\, G^A_{\mu\nu} G^{A\mu\nu}$ \\
$Q_{H\widetilde G}$         & $H^\dag H\, \widetilde G^A_{\mu\nu} G^{A\mu\nu}$ \\
$Q_{H W}$     & $H^\dag H\, W^I_{\mu\nu} W^{I\mu\nu}$ \\
$Q_{H\widetilde W}$         & $H^\dag H\, \widetilde W^I_{\mu\nu} W^{I\mu\nu}$ \\
$Q_{H B}$     & $ H^\dag H\, B_{\mu\nu} B^{\mu\nu}$ \\
$Q_{H\widetilde B}$         & $H^\dag H\, \widetilde B_{\mu\nu} B^{\mu\nu}$ \\
$Q_{H WB}$     & $ H^\dag \tau^I H\, W^I_{\mu\nu} B^{\mu\nu}$ \\
$Q_{H\widetilde W B}$         & $H^\dag \tau^I H\, \widetilde W^I_{\mu\nu} B^{\mu\nu}$ 
\end{tabular}
\end{minipage}
\begin{minipage}[t]{5.2cm}
\renewcommand{\arraystretch}{1.5}
\begin{tabular}[t]{c|c}
\multicolumn{2}{c}{$6:\psi^2 XH+\hbox{h.c.}$} \\
\hline
$Q_{eW}$      & $(\bar l_p \sigma^{\mu\nu} e_r) \tau^I H W_{\mu\nu}^I$ \\
$Q_{eB}$        & $(\bar l_p \sigma^{\mu\nu} e_r) H B_{\mu\nu}$ \\
$Q_{uG}$        & $(\bar q_p \sigma^{\mu\nu} T^A u_r) \widetilde H \, G_{\mu\nu}^A$ \\
$Q_{uW}$        & $(\bar q_p \sigma^{\mu\nu} u_r) \tau^I \widetilde H \, W_{\mu\nu}^I$ \\
$Q_{uB}$        & $(\bar q_p \sigma^{\mu\nu} u_r) \widetilde H \, B_{\mu\nu}$ \\
$Q_{dG}$        & $(\bar q_p \sigma^{\mu\nu} T^A d_r) H\, G_{\mu\nu}^A$ \\
$Q_{dW}$         & $(\bar q_p \sigma^{\mu\nu} d_r) \tau^I H\, W_{\mu\nu}^I$ \\
$Q_{dB}$        & $(\bar q_p \sigma^{\mu\nu} d_r) H\, B_{\mu\nu}$ 
\end{tabular}
\end{minipage}
\begin{minipage}[t]{5.4cm}
\renewcommand{\arraystretch}{1.5}
\begin{tabular}[t]{c|c}
\multicolumn{2}{c}{$7:\psi^2H^2 D$} \\
\hline
$Q_{H l}^{(1)}$      & $(H^\dag i\overleftrightarrow{D}_\mu H)(\bar l_p \gamma^\mu l_r)$\\
$Q_{H l}^{(3)}$      & $(H^\dag i\overleftrightarrow{D}^I_\mu H)(\bar l_p \tau^I \gamma^\mu l_r)$\\
$Q_{H e}$            & $(H^\dag i\overleftrightarrow{D}_\mu H)(\bar e_p \gamma^\mu e_r)$\\
$Q_{H q}^{(1)}$      & $(H^\dag i\overleftrightarrow{D}_\mu H)(\bar q_p \gamma^\mu q_r)$\\
$Q_{H q}^{(3)}$      & $(H^\dag i\overleftrightarrow{D}^I_\mu H)(\bar q_p \tau^I \gamma^\mu q_r)$\\
$Q_{H u}$            & $(H^\dag i\overleftrightarrow{D}_\mu H)(\bar u_p \gamma^\mu u_r)$\\
$Q_{H d}$            & $(H^\dag i\overleftrightarrow{D}_\mu H)(\bar d_p \gamma^\mu d_r)$\\
$Q_{H u d}$ + h.c.   & $i(\widetilde H ^\dag D_\mu H)(\bar u_p \gamma^\mu d_r)$\\
\end{tabular}
\end{minipage}

\vspace{0.25cm}

\begin{minipage}[t]{4.75cm}
\renewcommand{\arraystretch}{1.5}
\begin{tabular}[t]{c|c}
\multicolumn{2}{c}{$8:(\bar LL)(\bar LL)$} \\
\hline
$Q_{ll}$        & $(\bar l_p \gamma_\mu l_r)(\bar l_s \gamma^\mu l_t)$ \\
$Q_{qq}^{(1)}$  & $(\bar q_p \gamma_\mu q_r)(\bar q_s \gamma^\mu q_t)$ \\
$Q_{qq}^{(3)}$  & $(\bar q_p \gamma_\mu \tau^I q_r)(\bar q_s \gamma^\mu \tau^I q_t)$ \\
$Q_{lq}^{(1)}$                & $(\bar l_p \gamma_\mu l_r)(\bar q_s \gamma^\mu q_t)$ \\
$Q_{lq}^{(3)}$                & $(\bar l_p \gamma_\mu \tau^I l_r)(\bar q_s \gamma^\mu \tau^I q_t)$ 
\end{tabular}
\end{minipage}
\begin{minipage}[t]{5.25cm}
\renewcommand{\arraystretch}{1.5}
\begin{tabular}[t]{c|c}
\multicolumn{2}{c}{$8:(\bar RR)(\bar RR)$} \\
\hline
$Q_{ee}$               & $(\bar e_p \gamma_\mu e_r)(\bar e_s \gamma^\mu e_t)$ \\
$Q_{uu}$        & $(\bar u_p \gamma_\mu u_r)(\bar u_s \gamma^\mu u_t)$ \\
$Q_{dd}$        & $(\bar d_p \gamma_\mu d_r)(\bar d_s \gamma^\mu d_t)$ \\
$Q_{eu}$                      & $(\bar e_p \gamma_\mu e_r)(\bar u_s \gamma^\mu u_t)$ \\
$Q_{ed}$                      & $(\bar e_p \gamma_\mu e_r)(\bar d_s\gamma^\mu d_t)$ \\
$Q_{ud}^{(1)}$                & $(\bar u_p \gamma_\mu u_r)(\bar d_s \gamma^\mu d_t)$ \\
$Q_{ud}^{(8)}$                & $(\bar u_p \gamma_\mu T^A u_r)(\bar d_s \gamma^\mu T^A d_t)$ \\
\end{tabular}
\end{minipage}
\begin{minipage}[t]{4.75cm}
\renewcommand{\arraystretch}{1.5}
\begin{tabular}[t]{c|c}
\multicolumn{2}{c}{$8:(\bar LL)(\bar RR)$} \\
\hline
$Q_{le}$               & $(\bar l_p \gamma_\mu l_r)(\bar e_s \gamma^\mu e_t)$ \\
$Q_{lu}$               & $(\bar l_p \gamma_\mu l_r)(\bar u_s \gamma^\mu u_t)$ \\
$Q_{ld}$               & $(\bar l_p \gamma_\mu l_r)(\bar d_s \gamma^\mu d_t)$ \\
$Q_{qe}$               & $(\bar q_p \gamma_\mu q_r)(\bar e_s \gamma^\mu e_t)$ \\
$Q_{qu}^{(1)}$         & $(\bar q_p \gamma_\mu q_r)(\bar u_s \gamma^\mu u_t)$ \\ 
$Q_{qu}^{(8)}$         & $(\bar q_p \gamma_\mu T^A q_r)(\bar u_s \gamma^\mu T^A u_t)$ \\ 
$Q_{qd}^{(1)}$ & $(\bar q_p \gamma_\mu q_r)(\bar d_s \gamma^\mu d_t)$ \\
$Q_{qd}^{(8)}$ & $(\bar q_p \gamma_\mu T^A q_r)(\bar d_s \gamma^\mu T^A d_t)$\\
\end{tabular}
\end{minipage}

\vspace{0.25cm}

\begin{minipage}[t]{3.75cm}
\renewcommand{\arraystretch}{1.5}
\begin{tabular}[t]{c|c}
\multicolumn{2}{c}{$8:(\bar LR)(\bar RL)+\hbox{h.c.}$} \\
\hline
$Q_{ledq}$ & $(\bar l_p^j e_r)(\bar d_s q_{tj})$ 
\end{tabular}
\end{minipage}
\begin{minipage}[t]{5.5cm}
\renewcommand{\arraystretch}{1.5}
\begin{tabular}[t]{c|c}
\multicolumn{2}{c}{$8:(\bar LR)(\bar L R)+\hbox{h.c.}$} \\
\hline
$Q_{quqd}^{(1)}$ & $(\bar q_p^j u_r) \epsilon_{jk} (\bar q_s^k d_t)$ \\
$Q_{quqd}^{(8)}$ & $(\bar q_p^j T^A u_r) \epsilon_{jk} (\bar q_s^k T^A d_t)$ \\
$Q_{lequ}^{(1)}$ & $(\bar l_p^j e_r) \epsilon_{jk} (\bar q_s^k u_t)$ \\
$Q_{lequ}^{(3)}$ & $(\bar l_p^j \sigma_{\mu\nu} e_r) \epsilon_{jk} (\bar q_s^k \sigma^{\mu\nu} u_t)$
\end{tabular}
\end{minipage}
\end{center}
\caption{\label{op59}
The 59 independent dimension-six operators built from Standard Model fields which conserve baryon number, as given in 
Ref.~\cite{Grzadkowski:2010es}. The flavour labels of the form $p,r,s,t$ on the $Q$ operators are suppressed on the left hand side of
the tables.}
\end{table}


\bibliographystyle{JHEP}
\bibliography{RG}

\end{document}